\documentclass{article}
\usepackage{amsmath, amsthm, amssymb}
\newtheorem{lemma}{Lemma}
\newtheorem{thm}{Theorem}
\newtheorem{coro}{Corollary}

\def\<{\langle}
\def\>{\rangle}
\def\Ae{\mathcal A}
\def\Ce{\mathcal C}
\def\Be{\mathcal B}
\def\De{\mathcal D}
\def\Tr{{\rm Tr}}
\def\qed{{\hfill $\square$}}

\title{The structure of strongly additive states and Markov triplets on the CAR algebra}

\author{Anna Jen\v cov\'a \thanks{Supported by the grants VEGA 2/0032/09 and
meta-QUTE ITMS 26240120022}\\
{\small Mathematical Institute, Slovak Academy of Sciences,}\\
{\small \v {S}tef\'{a}nikova 49, 814 73 Bratislava, Slovakia} \\
{\small  jenca@mat.savba.sk} }

\date{}

\begin{document}

\maketitle
\abstract{We find a characterization  of states satisfying 
equality in strong subadditivity of entropy and of Markov triplets on the
CAR algebra. For even states, a more detailed structure of the density matrix  
is given.}

\section{Introduction}

A remarkable property of von Neumann entropy is the strong subadditivity (SSA): For a state $\rho$
on the 3-fold tensor product $B(\mathcal H_A\otimes\mathcal H_B\otimes \mathcal H_C)$, we have
\[
S(\rho)+S(\rho_B)\le S(\rho_{AB})+S(\rho_{BC})
\]
Here $\mathcal H_A$, $\mathcal H_B$ and $\mathcal H_C$ are finite dimensional Hilbert spaces and
$\rho_B$, $\rho_{AB}$, $\rho_{BC}$ are the restrictions of $\rho$ to the respective subsystems. 
This was first proved by Lieb and Ruskai in \cite{liebruskai}.

The structure of states that saturate the strong subadditivity of entropy, called strongly 
additive states,  was studied in \cite{hjpw}. In was shown that a state $\rho$ is strongly additive
if and only if it has the form
\begin{equation}\label{eq:ssaeq_hrpw}
\rho=\bigoplus_n A_n\otimes B_n,
\end{equation}
where $A_n\in B(\mathcal H_A\otimes \mathcal H_n)$ and $B_n\in B(\mathcal K_n\otimes \mathcal H_C)$ are positive operators
and $\mathcal H_B$ has a decomposition $\mathcal H_B=\bigoplus_n \mathcal H_n\otimes\mathcal K_n$ (see also \cite{japetz}, where
this was proved also for the infinite  dimensional case). Equivalently, 
\begin{equation} \label{eq:ssaeq_ja}
\rho= (D_{AB}\otimes I_C)(I_A\otimes D_{BC})
\end{equation}
where $D_{AB}\in B(\mathcal H_A\otimes \mathcal H_B)$ and $D_{BC}\in B(\mathcal H_B\otimes \mathcal H_C)$ are positive
matrices.

The Markov property for states in the quantum (non-commutative) probability was introduced by Accardi \cite{accardi} and Accardi and Frigerio \cite{acfrig},
in terms of completely positive unital maps, so-called quasiconditional expectations. For tensor products, it was shown 
that the Markov property is equivalent to strong additivity of the states \cite{ohyapetz}. 

The definition of the Markov property does not require the tensor product structure and can be applied in much more general 
situations. We are interested in the case of CAR algebras. The Markov states for 
CAR algebras were studied in \cite{afimu}. The strong subadditivity of entropy on 
CAR systems was recently shown and it was proved that strong additivity is 
equivalent to Markov property in the case of even states, see \cite{moriya}.
For noneven states, a necessary and sufficient condition for equality in (SSA) was given in \cite{belpit}.

The aim of the present paper is to find the structure of strongly additive states  and Markov triplets on the CAR algebra. We find an analogue of 
(\ref{eq:ssaeq_ja}) for any states and of (\ref{eq:ssaeq_hrpw}) for even states. This is done by a similar method as in \cite{japetz}, using the results of the theory of 
sufficient subalgebras.

The paper is organized as follows. The preliminary section summarizes the most important results on the CAR algebra and on sufficient subalgebras. The main tool used in the sequel is the factorization Theorem \ref{thm:factorization} in Section 2.1. Section 3 shows the relation between strong additivity and Markov property for any states on the CAR algebra. Section 4 contains the main results.

\section{Preliminaries}

\subsection{Sufficient subalgebras}

We first recall the definition and some characterizations of a sufficient 
subalgebra, which is a generalization of the classical notion of a sufficient statistic, see \cite{petz,ohyapetz} for details.

Let $\Ae$ be a finite dimensional algebra and let $\varphi,\psi$ be states on $\Ae$. Let $\Be\subset \Ae$ be a subalgebra and let $\varphi_0$, $\psi_0$ be the 
restrictions of the states to $\Be$. Then $\Be$ is sufficient for $\{\varphi,\psi\}$ is there is a completely positive, identity preserving map $E:\Ae\to \Be$, such that $\varphi_0\circ E=\varphi$, $\psi_0\circ E=\psi$.

For simplicity, let us further assume that the states are faithful. Let $\rho_\varphi$, $\rho_\psi$ be the densities of $\varphi$, $\psi$ with respect to a trace $\Tr$:
\[
\varphi(a)=\Tr \rho_\varphi a, \quad \psi(a)=\Tr \rho_\psi a,\qquad a\in \Ae
\]
The relative entropy $S(\varphi,\psi)$ is defined as
\[
S(\varphi,\psi)=S(\rho_\varphi,\rho_\psi)=\Tr \rho_\varphi(\log\rho_\varphi-\log \rho_\psi)
\]
It is monotone, in the sense that we have $S(\varphi,\psi)\ge S(\varphi_0,\psi_0)$ for any subalgebra $\Be\subseteq \Ae$. We will also need the definition of the 
generalized conditional expectation $E_\psi: \Ae\to \Be$ with respect to the state $\psi$ \cite{acccec}
\[
E_\psi(a)=E_{\rho_\psi}(a)=\rho_{\psi_0}^{-1/2}E_\Be(\rho_\psi^{1/2}a\rho_\psi^{1/2})\rho_{\psi_0}^{-1/2}
\]
where $E_\Be:\Ae \to \Be$ is the trace preserving conditional expectation. Then $E_\psi$ is a completely positive identity preserving map, such that 
$\psi_0\circ E_\psi=\psi$ and it is a conditional expectation if and only if 
$\rho^{it}_\psi\Be\rho^{-it}_\psi\subseteq \Be$ for all $t\in \mathbb R$.

The following theorem gives several equivalent characterizations of sufficiency.

\begin{thm}\label{thm:sufficiency} \cite{ohyapetz} The following conditions are 
equivalent.
\begin{enumerate}
\item[(i)] The subalgebra $\Be$ is sufficient for $\{\varphi,\psi\}$.
\item [(ii)] $S(\varphi,\psi)=S(\varphi_0,\psi_0)$.
\item[(iii)] $\rho_\varphi^{it}\rho_\psi^{-it}\in \Be$, for all $t\in \mathbb R$.
\item[(iv)] $E_\varphi=E_\psi$.
\end{enumerate}

\end{thm}

Our results below are based on the following generalization of the classical 
factorization criterion for sufficient statistics.

\begin{thm}\label{thm:factorization} \cite{japetz} Let $\varphi$, $\psi$ be
faithful states on $\Ae$ and let $\Be\subseteq \Ae$ be a subalgebra, such that 
$\rho_\psi^{it}\Be\rho_\psi^{-it}\subseteq \Be$ for all $t\in \mathbb R$. Then 
$\Be$ is sufficient for $\{\varphi,\psi\}$ if and only if 
\[
\rho_\varphi=\rho_{\varphi_0}D,\qquad \rho_\psi=\rho_{\psi_0}D
\]
where $\varphi_0=\varphi|_\Be$, $\psi_0=\psi|_\Be$ and $D$ is a positive element
in the relative commutant $\Be'\cap \Ae$.
\end{thm}

\subsection{The CAR    algebra}

We recall some basic facts about the CAR algebra, for details see \cite{armo, bratrob}.

The CAR algebra $\mathcal A$ is
the $C^*$- algebra generated by elements $\{a_i, i\in \mathbb Z\}$, 
satisfying the  anticommutation relations
\begin{equation}\label{eq:car}
a_ia_j+a_ja_i =0,\quad   a_ia_j^*+a_j^*a_i=\delta_{ij},\qquad i,j\in \mathbb Z
\end{equation}
For a subset  $I\subset \mathbb Z$, the $C^*$-subalgebra generated by 
$\{a_i, i\in I\}$  is denoted by $\mathcal A(I)$. If $I$ is finite, 
$\mathcal A(I)$ is isomorphic  to the full
matrix algebra $M_{2^{|I|}}(\mathbb C)$ by the so-called Jordan-Wigner 
isomorphism. Since 
\[
\mathcal A=\overline{\bigcup_{|I|<\infty}\mathcal A(I)}^{\, C^*},
\]
there is a unique tracial state $\tau$ on $\mathcal A$, obtained as an 
extension of the unique tracial  states on $\mathcal A(I)$, $|I|<\infty$. It has 
the following product property:
\begin{equation}\label{eq:product}
\tau(ab)=\tau(a)\tau(b),\qquad a\in \mathcal A(I),\ b\in \mathcal A(J),\quad I\cap J=\emptyset
\end{equation}

\subsubsection{Graded commutation relations}

For $I\subseteq \mathbb Z$, we denote by $\Theta^I$ the (unique)
automorphism of $\mathcal A$, such that
\begin{equation}\label{eq:thetaI}
\Theta^I(a_i)=-a_i, \ i\in I,\qquad \Theta^I(a_i)=a_i,\ i\notin I
\end{equation}
in particular, we denote $\Theta^{\mathbb Z}$ by $\Theta$. The even  and odd 
parts  of $\mathcal A$ are defined as 
\[
\mathcal A_+:=\{ a\in \mathcal A,\ \Theta(a)=a\},\ 
\mathcal A_-:=\{ a\in \mathcal A,\ \Theta(a)=-a\}
\]
and  $\mathcal A(I)_+:=\mathcal A(I)\cap \mathcal A_+$, $\mathcal A(I)_-:=
\mathcal A(I)\cap \mathcal A_-$. Let  $I\cap J=\emptyset$
 and $a\in \mathcal A(I)_\sigma$, $b\in \mathcal A(J)_{\sigma'}$, 
 $\sigma,\sigma'\in \{+,-\}$. Then we have the graded commutation 
relations
\begin{equation}\label{eq:graded}
ab=\epsilon(\sigma,\sigma')ba
\end{equation}
where
\begin{eqnarray*}
\epsilon(\sigma,\sigma')&=& -1\quad  \mbox{if } \sigma=\sigma'=-\\
			&=& +1\quad  \mbox{otherwise} 
\end{eqnarray*}

If $I$ is finite,
 then there is a self-adjoint unitary $v_I\in \mathcal A(I)$, such that 
 $\Theta^I(a)=v_Iav_I$ for $a\in \mathcal A$ and
 \begin{equation}\label{eq:vI}
v_I=\Pi_{i\in I}v_i,\qquad v_i=a_i^*a_i-a_ia_i^*
 \end{equation}
Note that $v_iv_j=v_jv_i$ if $i\neq j$ and $\tau(v_i)=0$. Moreover,
$v_I\in \mathcal A(I)_+$ and $\mathcal A(I)_+=\mathcal A_I\cap \{v_I\}'$.

\subsubsection{Matrix units}

Let $A\subset \mathbb Z$ be a finite set, $A=\{i_!,\dots,i_n\}$. The relations
\begin{eqnarray*}
e_{11}^{(i_j)}:=a_{i_j}a_{i_j}^*,\qquad e_{12}^{(i_j)}:=V_{i_{j-1}}a_{i_j}\\
e^{(i_j)}_{21}:=V_{i_{j-1}}a_{i_j}^*,\qquad e_{22}^{(i_j)}:=a_{i_j}^*a_{i_j}
\end{eqnarray*}
with $V_{i_j}=\Pi_{k=1}^j(I-2a_{i_k}^*a_{i_k})$ define a family of mutually 
commuting $2\times 2$ matrix units. The Jordan-Wigner isomorphism is then given by\[
e^{(A)}_{k_1l_1\dots k_nl_n}:=e^{(i_1)}_{k_1l_1}\dots e^{(i_n)}_{k_nl_n}\mapsto
e_{k_1l_1}\otimes\dots\otimes e_{k_nl_n}
\]
where $e_{kl}$ are standard matrix units in $M_2(\mathbb C)$. The elements 
$\{e^{(a)}_\alpha,\alpha\in \mathcal J(A):=(\{1,2\}\times \{1,2\})^n\}$  span $\Ae(A)$. Note that $e^{(A)}_\alpha$ are either even or odd, we denote the set of 
indices of the even resp. odd elements by $\mathcal J(A)_+$, resp. 
$\mathcal J(A)_-$. Moreover, the elements $p_\alpha^{(A)}:=e_\alpha^{(A)}(e_\alpha^{(A)})^*$ and $q_\alpha^{(A)}:=(e^{(A)}_\alpha)^*e^{(A)}_\alpha$ are even 
projections in $\Ae(A)$ and 
\begin{equation}\label{eq:munits}
p_\alpha^{(A)}e^{(A)}_\beta q_\alpha^{(A)}=\delta_{\alpha,\beta}e^{(A)}_\alpha, 
\qquad \alpha,\beta \in \mathcal J(A)
\end{equation}

\subsubsection{Conditional expectations}

Let $I\subseteq \mathbb Z$ be any subset. Then there is a unique conditional 
expectation $E_I: \Ae\to \Ae (I)$, satisfying
\begin{equation}\label{eq:condexp}
\tau(ab)=\tau(E_I(a)b),\qquad a\in \Ae,\ b\in \Ae(I)
\end{equation}
This implies that $\Theta E_I=E_I\Theta$. If $J\subseteq \mathbb Z$, then 
$E_I(a)\in \Ae(I\cap J)$ for $a\in \Ae(J)$ and $E_IE_J=E_JE_I=E_{I\cap J}$. Note
also  that the product property (\ref{eq:product}) implies that for $a\in \Ae(J)$ with $I\cap J=\emptyset$, $E_I(a)=\tau(a)$.

\section{Strong additivity and Markov property}

Let $A$, $B$, $C$ be disjoint finite subsets in $\mathbb Z$. Let us denote $\Ae=\Ae_{ABC}=\Ae(A\cup B\cup C)$, $\Ae_{AB}=\Ae(A\cup B)$ etc. 
Let $\varphi$ be a faithful state on $\Ae$ and let $\rho$ be its density, that is,
$\varphi(x)=\Tr \rho x$ for $x\in \Ae$.

Let $\varphi_{AB}$ denote  the  restriction of $\varphi$ 
to $\Ae_{AB}$, similarly $\varphi_{BC}$ and $\varphi_B$. Then the density of $\varphi_{AB}$ in $\Ae_{AB}$ is 
\[
\rho_{AB}=E_{AB}(\rho),
\]
where $E_{AB}=E_{A\cup B}$. As an element in $\Ae$, $\rho_{AB}$ is the density of the state $\varphi\circ E_{AB}$.

\subsection{Strong subadditivity of entropy}

Let $\rho$ be the density of the state $\varphi$. Let
\[
S(\varphi)= -\Tr \rho(\log (\rho))
\]
be the von Neumann entropy of $\varphi$. The strong subadditivity for CAR 
algebras\begin{equation*} \tag{SSA}
S(\varphi)-S(\varphi_{AB})-S(\varphi_{BC})+S(\varphi_B)\le 0
\end{equation*}
 was proved in \cite{moriya}.
This inequality is equivalent with
\[
S(\rho,\rho_{BC})-S(\rho_{AB},\rho_B)\ge 0.
\]
Since $\rho_{AB}=E_{AB}(\rho)$, $\rho_B=E_{AB}(\rho_{BC})$ are restrictions of $\rho$ and $\rho_{BC}$ to $\Ae_{AB}$, this  holds by monotonicity of the relative 
entropy. Theorem \ref{thm:sufficiency} (ii)  then implies the  following.

\begin{thm}\label{thm:SSAsuff} The equality in (SSA) is attained if and only if the subalgebra 
$\Ae_{AB}$ is sufficient for $\{\varphi,\varphi\circ E_{BC}\}$.

\end{thm}

\subsection{Markov triplets and strong additivity}

The state $\varphi$ is a Markov triplet if there exists a completely positive, identity preserving map $E: \Ae \to \Ae_{AB}$, such that 
\begin{enumerate}
\item[(i)] $E(xy)=xE(y)$, for all $x\in \Ae_A$ and $y\in \Ae$.
\item[(ii)] $\varphi\circ E=\varphi$
\item[(iii)] $E(\Ae_{BC})\subseteq \Ae_B$
\end{enumerate}
The map $E$ is called a quasi-conditional expectation with respect to the triplet $\Ae_A \subset \Ae_{AB}\subset \Ae$.
Let us now define the subalgebras $\Be\subset \Ce$ in $\Ae_{AB}$ by
\[
\Ce=\{ x\in \Ae_{AB}, \rho_{BC}^{it}x\rho_{BC}^{-it}\in \Ae_{AB}\}, \qquad
\Be=\{ y\in \Ae_B, \rho_{BC}^{it}y\rho_{BC}^{-it}\in \Ae_B\}
\]
Note that $\Ce$ is the fixed point subalgebra of the generalized conditional 
expectation $E_{\rho_{BC}}:\Ae \to \Ae_{AB}$ with respect to $\rho_{BC}$ 
\cite{acccec}.
We also have $E_{BC}(\Ce)= \Be$. Indeed, if $x=E_{BC}(y)$ for
some $y\in \Ce$, then 
\[
\rho_{BC}^{it}x\rho_{BC}^{-it}=E_{BC}(\rho_{BC}^{it}y\rho_{BC}^{-it})\in \Ae_B,
\]
so that $E_{BC}(\Ce)\subseteq \Be$, the converse inclusion is clear.

\begin{thm}\label{thm:SSAmarkov} The state $\varphi$ is a Markov triplet if and only if $\varphi$
satisfies  equality in (SSA) and $\Ae_A\subseteq \Ce$.
\end{thm}

{\it Proof.} Let $\varphi$ be a Markov triplet and let $E$ be the 
quasi-conditional expectation. 
Then $E$ is a completely positive identity preserving map
$\Ae\to\Ae_{AB}$  and $\varphi \circ E=\varphi$. 
Moreover, let $x\in \Ae_A$, $y\in \Ae_{BC}$, then 
\[
\varphi\circ E_{BC}\circ E(xy)=\varphi\circ E_{BC}(xE(y))=\tau(x)\varphi(E(y))=
\tau(x)\varphi(y)=\varphi\circ E_{BC}(xy)
\]
Since by the commutation relations (\ref{eq:car}) $\Ae$ is spanned by elements of the form $xy$, the above equality implies that $E$ preserves 
$\varphi\circ E_{BC}$ as well, so that $\Ae_{AB}$ is sufficient
 for $\{\varphi,\varphi\circ E_{BC}\}$ and equality in (SSA) holds by Theorem \ref{thm:SSAsuff}. Let
 \[
F=\lim_n \frac 1n \sum_{k=0}^{n-1}E^k
 \]
By the ergodic theorem, $F$ is a conditional expectation with range $\mathcal R(F)$ the fixed point subalgebra of $E$. By the property (i) of Markov triplets,
$\Ae_A\subseteq \mathcal R(F)$. Since $F$ also preserves $\varphi\circ E_{BC}$, we have by Takesaki theorem that $\rho_{BC}^{it}\mathcal R(F) \rho_{BC}^{-it}\subseteq \mathcal R(F)$, hence also $\rho^{it}_{BC}\Ae_A\rho_{BC}^{-it}\subseteq \mathcal R(F)\subseteq \Ae_{AB}$. It follows that $\Ae_A\subseteq \Ce$.

Conversely, suppose equality in (SSA) and $\Ae_A\subseteq \Ce$. Let 
$E=E_{\rho_{BC}}:\Ae\to\Ae_{AB}$ be the generalized conditional expectation.
By Theorem \ref{thm:SSAsuff}, $\Ae_{AB}$ is sufficient for for $\{\varphi,\varphi\circ E_{BC}\}$ and by Theorem \ref{thm:sufficiency} (iv), $E_{\rho_{BC}}=E_\rho$,
 hence $\varphi\circ E=\varphi$. By the assumptions, $\Ae_A\subseteq \Ce$ the 
 fixed point subalgebra of $E$. The property (iii) of Markov triplets is clear 
 from the definition of $E_{\rho_{BC}}$.

\qed

The following Corollary was already proved in \cite{moriya}.

\begin{coro}\label{coro:even} Let $\varphi$ be an even state. Then $\varphi$ is a Markov triplet if and only if it satisfies equality in (SSA). 
\end{coro}

{\it Proof.} Since $\rho$ is even, $\rho_{BC}$ is even as well and we always have
$\Ae_A\subseteq \Ce$, by the graded  commutation relations. The proof now 
follows  from Theorem \ref{thm:SSAmarkov}.

\qed

\section{Characterization of strongly additive states and Markov triplets}

\begin{thm}\label{thm:SSAfactor} The state $\varphi$ satisfies equality in (SSA) if and only if there 
are positive elements $x\in \Ae_{AB}$, $y\in \Ae_{BC}$, such that 
\begin{equation}\label{eq:factor}
\rho=xy
\end{equation}

\end{thm}

{\it Proof.} Suppose that $\varphi$ satisfies equality in (SSA). Then $\Ae_{AB}$ 
is a sufficient subalgebra for $\{\varphi,\varphi\circ E_{BC}\}$. By Theorem \ref{thm:sufficiency}, this implies that $u_t:=\rho^{it}\rho_{BC}^{-it}\in \Ae_{AB}$ for all $t$.  Since
$\rho_{BC}^{it}u_s\rho_{BC}^{-it}=u_t^*u_{s+t}$ for $s,t\in \mathbb R$, this 
implies that $u_t\in \Ce$ for all $t$. Hence, $\Ce$ is a sufficient subalgebra as well,  such that $\rho_{BC}^{it}\Ce\rho_{BC}^{-it}\subseteq 
\Ce$. By Theorem \ref{thm:factorization},  
\begin{eqnarray*}
\rho&=&xy\\
\rho_{BC} &=& x_0y
\end{eqnarray*}
where $x,x_0\in \Ce\subseteq \Ae_{AB}$ are  the densities of
the restrictions $\varphi|_\Ce$ and $\varphi\circ E_{BC}|_\Ce$ and
$y$ is a positive element in $\Ce'$.  Note also that $\varphi\circ E_{BC}|_\Ce$ 
is 
the restriction of $\varphi$ to $E_{BC}(\Ce)=\Be\subseteq \Ae_B$, so that 
$x_0\in \Ae_B$.

By the graded commutation relations, we  have $(\Ae_A)_+ \subseteq \Ce$, so that 
$\Ce'\subseteq ((\Ae_A)_+)'= \Ae_{BC}+v_A\Ae_{BC}$, \cite{armo} (all commutants are taken in the algebra $\Ae$). 
Let $y\in \Ce'$, then  $y=d_1+v_Ad_2$, where $d_1,d_2\in \Ae_{BC}$. We have
\[
x_0y=E_{BC}(x_0y)=E_{BC}(x_0(d_1+v_Ad_2))=x_0d_1.
\]
Since $\varphi$, and therefore also its restriction to $\Be$ is faithful, $x_0$ is invertible, so that $y=d_1\in \Ae_{BC}$.

Conversely, suppose $\rho=xy$ as above. 
Then  $\rho_{AB}=xy_0$, $\rho_{BC}= x_0y$ and $\rho_B=x_0y_0$, 
where $y_0=E_{AB}(y)\in \Ae_B$, $x_0=E_{BC}(x)\in \Ae_B$. Clearly, both $x$ and $x_0$ must commute with both $y$ and $y_0$. Then 
$\rho^{it}\rho_{BC}^{-it}= x^{it}x_0^{-it}\in \Ae_{AB}$. By Theorem \ref{thm:sufficiency} (iii), $\Ae_{AB}$ is sufficient for $\{\varphi,\varphi\circ E_{BC}\}$, so that $\varphi$ satisfies equality in (SSA).

\qed

\begin{thm}\label{thm:Markovfactor} The state $\varphi$ is a Markov triplet if and only if there are
positive elements $x\in \Ae_{AB}$ and $y\in (\Ae_{BC})_+$, such that 
\[
\rho=xy
\]

\end{thm}

{\it Proof.} Let $\varphi$ be a Markov triplet. By Theorem \ref{thm:SSAmarkov},
 $\varphi$ satisfies equality in (SSA) and  by Theorem
 \ref{thm:SSAfactor} and its proof, there are positive elements $x\in \Ce$, $y\in \Ce'$, such 
 that  $\rho=xy$. Since  $\Ae_A\subseteq \Ce$, $\Ce'\subseteq \Ae_A'=(\Ae_{BC})_++v_A(\Ae_{BC})_-$, \cite{armo}. This implies that $y=d_++v_Ad_-$, where $d_+\in (\Ae_{BC})_+$ and
 $d_-\in (\Ae_{BC})_-$. By the same reasoning as in the proof of Theorem \ref{thm:SSAfactor}, we get that $y=d_+\in (\Ae_{BC})_+$.

Conversely, let $\rho=xy$ as above, then $\varphi$ satisfies equality in (SSA) by 
Theorem \ref{thm:SSAfactor}, and $\rho_{BC}=x_0y$, $x_0=E_{BC}(x)$. For $a\in \Ae_A$, 
\[
\rho_{BC}^{it}a\rho_{BC}^{-it} =x_0^{it}ax_0^{-it}\in \Ae_{AB}
\]
by the graded commutation relations, so that $\Ae_A\subseteq \Ce$. By Theorem 
\ref{thm:SSAmarkov}, $\varphi$ is a Markov triplet.

\qed

\subsection{Even Markov triplets}

\begin{thm}\label{thm:evenmarkovfactor}
Let $\varphi$ be an even state. Then $\varphi$ is a Markov triplet if and only if there are positive elements $x\in \Ae_{AB}$ and $y\in \Ae_{BC}$, such that 
\[
\rho=xy.
\]
Moreover, $x$ and $y$ can be chosen even.

\end{thm}

{\it Proof.} Follows easily from Corollary \ref{coro:even},
Theorems \ref{thm:SSAfactor} and \ref{thm:Markovfactor} and the fact that $\rho$ is even.

\qed

We will now describe the subalgebras $\Ce$ and $\Ce'$ for even states. 
Since $\rho_{BC}$ is even, both $\Ce$ and $\Be$  and  their  commutants 
$\Ce'$ and  $\Be'$ are invariant under $\Theta$.

\begin{lemma}\label{lemma:c} If $\varphi$ is even, then 
\[
\Ce= \Ae_A\bigvee \Be
\]
\end{lemma}

{\it Proof.} Since  $\Ae_A\subseteq \Ce$ and clearly 
also $\Be\subseteq \Ce$, we have  $\Ae_A\bigvee \Be\subseteq \Ce$.

Conversely, any element $x\in \Ce\subseteq \Ae_{AB}$ has the form 
$x=\sum_\alpha e^{(A)}_\alpha
b_\alpha$  for some $b_\alpha \in \Ae_B$, where $e^{(A)}_\alpha$ are the matrix 
units in  $\Ae_A$. By (\ref{eq:munits}), we have for any $\alpha$, 
\[
p^{(A)}_\alpha x q^{(A)}_\alpha=
e^{(A)}_\alpha b_\alpha,
\]
since $q^{(A)}_\alpha$ is always even. As $\Ae_A\subseteq \Ce$,   
this implies that 
$e^{(A)}_\alpha b_\alpha\in \Ce$ for all $\alpha$. It follows
 that 
 \[
 \rho^{it}_{BC}e^{(A)}_\alpha b_\alpha\rho^{-it}_{BC}=e^{(A)}_\alpha
 \rho_{BC}^{it}b_\alpha \rho_{BC}^{-it}\in \Ae_{AB},
 \] 
 hence $b_\alpha\in \Be$, so that $\Ce \subseteq  \Ae_A\bigvee \Be$.

\qed 

\begin{lemma}\label{lemma:ccom} If $\varphi$ is even, then 
\[
\Ce'=(\Be'\cap \Ae_{BC})_++(\Be'\cap \Ae_{BC})_- v_A
\]
\end{lemma}

{\it Proof.}  Since $\Ae_A\subseteq \Ce$, we have $\Ce'\subseteq \Ae_A'= 
(\Ae_{BC})_++ (\Ae_{BC})_-v_A$, by \cite{armo}. Let $d_++v_Ad_-\in \Ce'$ and 
let $x\in \Be\subset \Ce$. Then we must have 
$xd_+-d_+x=v_A(d_-x-xd_-)$. Applying $E_{BC}$ on both sides, we get 
$xd_+-d_+x=d_-x-xd_-=0$, hence $d_+\in \Be'\cap (\Ae_{BC})_+=(\Be'\cap \Ae_{BC})_+$, $d_-\in \Be'\cap (\Ae_{BC})_-=(\Be'\cap \Ae_{BC})_-$. 

Conversely, let $d_+ \in (\Be'\cap \Ae_{BC})_+$, $d_-\in  (\Be'\cap \Ae_{BC})_-$
and let $a\in \Ae_A$, $b\in \Be$. Then by the graded commutation relations,
\[
ab(d_++v_Ad_-)=d_+ab+ v_Ad_-a_+b+v_Ad_-a_-b=(d_++v_Ad_-)ab
\]
so that $d_++v_Ad_-\in \Ce'$.
\qed

\begin{lemma}\label{lemma:b}
Denote 
$\tilde {\Be} = \Be'\cap \Ae_B$. Then  
\[
\Be'\cap \Ae_{BC}=\tilde {\Be}\bigvee ((\Ae_C)_++v_B(\Ae_C)_-)
\]

\end{lemma}

{\it Proof.} It is easy to see that both 
$\tilde {\Be}$ and  $(\Ae_C)_++v_B(\Ae_C)_-$ are subsets in $\Be'\cap \Ae_{BC}$. 
Conversely,
any $y\in \Ae_{BC}$ has the form 
$y=\sum_\beta b_\beta e^{(C)}_\beta$, for some $b_\beta\in \Ae_B$. Let $x\in \Be$,
then 
\begin{eqnarray*}
yx&=&\sum_\beta b_\beta e^{(C)}_\beta(x_++x_-)=\sum_\beta b_\beta x_+e^{C}_\beta+
\sum_{\beta\in \mathcal J(C)_+}b_\beta x_-e^{(C)}_\beta -
\sum_{\beta\in \mathcal J(C)_-}b_\beta x_-e^{(C)}_\beta\\
&=& \sum_{\beta\in \mathcal J(C)_+} b_\beta xe^{(C)}_\beta + 
\sum_{\beta\in \mathcal J(C)_-}b_\beta\Theta(x)
e^{(C)}_\beta
\end{eqnarray*}
It follows that $yx=xy$  only if 
$xb_\beta=b_\beta x$ for  $\beta\in \mathcal J(C)_+$ and $xb_\beta=b_\beta\Theta(x)$ for $\beta\in \mathcal J(C)_-$. This is true for all $x\in \Be$ if and only if $b_\beta\in \Be'$ for
$\beta\in \mathcal J(C)_+$ and $b_\beta v_B\in \Be'$ for $\beta\in \mathcal J(C)_-$, this implies the statement of the lemma.

\qed

Let us now look at the algebra $\Be$. Let $P_1,\dots, P_m$ be the minimal central projections in $\Be$. Since  $\Be$ is  invariant under $\Theta$, we must have for each $i$, $\Theta(P_i)=P_j$ for some $j$. Suppose that 
$\Theta(P_i)=P_i$, $i=1,\dots,k$ and $\Theta(P_i)=P_{i+1}$ for $i=k+2l+1$, $l=0,\dots, \frac {m-k}2-1$. 

\begin{lemma} \label{lemma:central} Let us denote $P_A=\frac12 (1+v_A)$. 
The minimal central projections in $\Ce$ are
\begin{eqnarray*}
Q_i&:=&P_i, \qquad i=1,\dots,k\\
Q_{k+1}&:=& P_AP_{k+1}+(1-P_A)P_{k+2}, \quad Q_{k+2}:=(1-P_A)P_{k+1}+P_AP_{k+2}\\
\dots\\
Q_{m-1}&:=& P_AP_{m-1}+(1-P_A)P_{m},\quad Q_m:=(1-P_A)P_{m-1}+P_AP_{m}
\end{eqnarray*}

\end{lemma}

{\it Proof.} Clearly,
$\mathcal Z(\Ce)\subset \Ae_A'\cap \Ae_{AB}=(\Ae_B)_++v_A(\Ae_B)_-$ and it is easy to see that if $x_++v_Ax_-\in \mathcal Z(\Ce)$, then $x_+,x_-$ must be in $\mathcal Z(\Be)$. Therefore, $x_+=\sum c_j P_j$ and $x_-=\sum_j d_jP_j$, for some 
$c_j,d_j\in \mathbb C$. Since $x_+$ is even, we must have 
$c_j=c_{j+1}$ for $j=k+2l+1$, $l=0,\dots,\frac{m-k}2-1$. Similarly, we get
$d_j=0$ for $j=1,\dots,k$ and $d_j=-d_{j+1}$ for $j=k+2l+1$, $l=0,\dots,\frac{m-k}2-1$.

Suppose now that $P=x_++v_Ax_-$ is a projection, then 
we must have $x_+^*x_++x_-^*x_-=x_+$ and $x^*_+x_-+x_-^*x_+=x_-$. This implies 
that $c_j=|c_j|^2$ for $j=1,\dots,k$, $c_j=|c_j|^2+|d_j|^2\ge0$ and $c_j(d_j+\bar d_j)=d_j\in \mathbb R$, for $j>k$.
Hence $2c_jd_j=d_j$, so that either $d_j=0$ and then $c_j=c_j^2$, or $c_j=\frac12$ and then $d_j=\pm 
\frac12$.

It follows that any projection in $\mathcal Z(\Ce)$ is a sum of some of 
the following  projections: $P_i$, $i=1,\dots,k$, $P_j+P_{j+1}$, $j=k+2l+1$, and 
$\frac 12 (P_j+P_{j+1}\pm v_A(P_j-P_{j-1}))$, $j=k+2l+1$. Since the last 
projection is equal to $Q_j$ or $Q_{j+1}$ and $Q_j+Q_{j+1}=P_j+P_{j+1}$, the 
Lemma follows.

\qed

\begin{thm} Let  $\varphi$ be an even faithful state on $\Ae$. Then $\varphi$ 
is a Markov triplet if and only if there is an orthogonal family of 
projections 
$P_1,\dots,P_m\in \Ae_B$ and  decompositions 
$P_j\Ae_BP_j=\Be_j\otimes \tilde \Be_j$, where $\Be_j$ and $\tilde \Be_j$ are 
full matrix algebras,  such that 
\begin{enumerate}
\item $\Theta(P_j)=P_j$ and $\Be_j$ and $\tilde \Be_j$ are invariant under $\Theta$ for
$j=1,\dots,k$ 
\item $\Theta(P_j)=P_{j+1}$ and $\Theta(\Be_j)=\Be_{j+1}$, $\Theta(\tilde \Be_j)=
\tilde \Be_{j+1}$ for $j=k+2l+1$, $l=0,\dots,\frac{m-k}2-1$
\item Let us denote 
\begin{eqnarray*}
 V_j&=&P_jv_B\\
 \Ce_j &=& \Ae_A\bigvee \Be_j\\
 \tilde \Ce_j &=& \tilde\Be_j\bigvee ((\Ae_C)_++V_j(\Ae_C)_-)
\end{eqnarray*}
  for $j=0,\dots,k$ and 
\begin{eqnarray*}  
  U_l&=&(P_{k+2l+1}+P_{k+2l+2})v_B\\
  \De_l &=& \Ae_A\bigvee (P_A\Be_{k+2l+1}+(1-P_A) \Be_{k+2l+2})\\
  \tilde \De_l &=&  (P_A\tilde \Be_{k+2l+1}+(1-P_A)\tilde \Be_{k+2l+2})\bigvee ((\Ae_C)_+ +U_l(\Ae_C)_-)
\end{eqnarray*}

  for $l=0,\dots,\frac{m-k}2-1$, then there is a decomposition 
\begin{equation} \label{eq:evendecomp}
\rho=\bigoplus_{j=1}^k x_j\otimes y_j\oplus \bigoplus_{l=0}^{\frac{m-k}2-1}
(z_l\otimes w_l\oplus \Theta(z_l\otimes w_l)),
\end{equation}
where $x_j\in \Ce_j$ and
$y_j\in \tilde\Ce_j$  are positive and even  
for $j=1,\dots,k$, and $z_l\in \De_l$,  $w_l\in \tilde \De_l$ are positive for 
$l=0,\dots,\frac{m-k}2-1$.  

\end{enumerate}

\end{thm}

{\it Proof.} Suppose that  $\rho$ has the form (\ref{eq:evendecomp}). Let us 
define  $Q_1,\dots, Q_m$ from $P_1,\dots,P_m$
as in Lemma \ref{lemma:central}. Then $Q_j$ are mutually orthogonal projections 
and it is easy to see that $Q_jx_j=x_j$, $Q_jy_j=y_j$ and
$Q_{k+2l+1}z_l=z_l$, $Q_{k+2l+1}w_l=w_l$, $Q_{k+2l+2}\Theta(z_l)=\Theta(z_l)$,
$Q_{k+2l+2}\Theta(w_l)=\Theta(w_l)$. Put
\[
x=\bigoplus_j x_j\oplus \bigoplus_l (z_l\oplus \Theta(z_l)), \qquad 
y=\bigoplus_j y_j\oplus \bigoplus_l (w_l\oplus \Theta(w_l))
\]
then $x\in \Ae_{AB}$ and $y\in \Ae_{BC}$ are positive even elements and 
$\rho=xy$. By Theorem \ref{thm:evenmarkovfactor}, this implies that $\varphi$ is a Markov triplet.

Conversely, suppose that $\varphi$ is an even Markov triplet. Then we have seen 
that $\rho=xy$, where $x\in \Ce$ and $y\in \Ce'$ are positive and even. By Lemmas
 \ref{lemma:c}, \ref{lemma:ccom} and \ref{lemma:b}, $x\in \Ae_A\bigvee \Be$ and
 $y\in \tilde \Ce:= \tilde \Be\bigvee ((\Ae_C)_++v_B (\Ae_C)_-)$.

Let $P_1,\dots,P_m$ be the minimal central projections in $\Be$ and let 
$\Be_j:= P_j\Be$, $\tilde \Be_j:=P_j\tilde \Be$.  Then $\Be_j$ and $\tilde \Be_j$ are full matrix algebras and $P_j \Ae_BP_j=\Be_j\otimes \tilde \Be_j$. Moreover,
 we may suppose that there is some $k\le m$ such that 1. and 2. are fulfilled.

The minimal central projections $Q_1,\dots,Q_m$  in $\Ce$
are given by Lemma \ref{lemma:central}. 
Let us denote $\Ce_j=Q_j\Ce_j$, $\Ce'_j =Q_j\Ce'$.
Then each $\Ce_j$, $\Ce'_j$ is isomorphic to a full matrix 
algebra and we have a decomposition
\[
\Ce=\bigoplus_j \Ce_j\otimes \tilde I_j,\qquad \Ce'=\bigoplus_j I_j\otimes 
\Ce'_j 
\]
Since we are interested only in even elements in $\Ce'$, we take the algebra
$\tilde \Ce_j:=Q_j\tilde \Ce\subset \Ce'_j$.
For $j=1,\dots,k$, $\Ce_j$ are invariant under $\Theta$. For $l=0,\dots, \tfrac{m-k}2-1$, let us denote 
$\De_l:=\Ce_{k+2l+1}$, $E_l:= Q_{k+2l+1}+Q_{k+2l+2}=P_{k+2l+1}+P_{k+2l+2}$. Then 
$E_l$ is an even projection, the algebra $E_l\Ce= \De_l\oplus \Theta(\De_l)$ is 
invariant under $\Theta$ and even elements in $E_l\Ce$ are of the form 
$x\oplus \Theta(x)$, for some $x\in \De_l$. Similar relation hold for $\Tilde \Ce_j$ and $\tilde \De_l:=\tilde \Ce_{k+2l+1}$.

Let us denote $x_j:=Q_jx$, $y_j:=Q_jy$ for $j=1,\dots,k$ and
$x_j:=E_lx$, $y_j:=E_ly$ for $j=k+2l+1$, $l=0,\dots, \frac{m-k}2-1$. 
Then all $x_j$, $y_j$ are positive and even and
\[
\rho=\bigoplus_{j=1}^k xj\otimes y_j \oplus \bigoplus_{l=0}^{\frac{m-k}2-1}
x_{k+2l+1}y_{k+2l+1}
\]
Moreover,  for $j=k+2l+1$ we must have 
$x_j=z_l\oplus \Theta(z_l)$ for some positive $z_l\in \De_l$ and similarly 
$y_j=w_l\oplus \Theta(w_l)$ for positive $w_l\in \tilde \De_l$, $l=0,\dots,\frac{m-k}2-1$. The rest of the proof now follows from Lemmas \ref{lemma:c} and \ref{lemma:b}.

\qed


\begin{thebibliography}{99}

\bibitem{accardi} L. Accardi, On noncommutative Markov property, Funct. Anal. 
Appl. {\bf 9} (1975), 1-8

\bibitem{acccec} L. Accardi, C. Cecchini, Conditional expectations in von Neumann algebras and a theorem of Takesaki, J. Functional. Anal. {\bf 45}(1982), 245-273.


\bibitem{acfrig} L. Accardi and A. Frigerio, Markovian cocycles, Math. Proc. R. Ir. Acad. {\bf 83} (1983), 251-263

\bibitem{afimu} L. Accardi, F. Fidaleo, F. Mukhamedov, 
Markov states and chains on the CAR algebra, Inf. Dimen. Anal. Quantum Probab., 
Rel. Top. {\bf 10} (2007), 165-184 

\bibitem{armo} H. Araki, H. Moriya, Equilibrium statistical mechanics of fermion lattice systems, Rev. Math. Phys. {\bf 15} (2003), 93-198 

\bibitem{belpit} J. Pitrik, V.P. Belavkin, Notes on the equality in SSA of entropy on CAR algebra, http://arxiv.org/abs/math-ph/0602035, (2006)

\bibitem{bratrob} O. Bratelli and D.W. Robinson, {\it Operator Algebras and 
Quantum Statistical Mechanics II}, Springer-Verlag, Heidelberg, 1981

\bibitem{hjpw} P. Hayden, R. Jozsa, D. Petz, A. Winter, Structure of states which satisfy strong subadditivity of quantum entropy with equality, Commun. Math. Phys.
{\bf 246} (2004), 359-374

\bibitem{japetz} A. Jen\v cov\'a and D. Petz, Sufficiency in quantum statistical 
inference, Commun. Math. Phys. {\bf 263}(2006), 259-276. 

\bibitem{liebruskai} E.H. Lieb and M.B. Ruskai, Proof of the strong subadditivity of quantum-mechanical entropy, J. Math. Phys. {\bf 14} (1973), 1938-1941

\bibitem{moriya} H. Moriya, Markov property and strong additivity of von Neumann 
entropy for graded systems, J. Math. Phys. {\bf 47} (2006), 033510 

\bibitem{ohyapetz} M. Ohya and D. Petz, {\it Quantum Entropy and Its Use}, Springer-Verlag, Heidelberg, 1993

\bibitem{petz} D. Petz, Sufficiency of channels over von Neumann algebras, Quart. J. Math. Oxford {\bf 39} (1988), 97-108


\end{thebibliography}
\end{document}